\documentclass[prd,amsmath,amssymb,preprintnumbers,superscriptaddress,twocolumn,10pt,nofootinbib,showkeys,showpacs]{revtex4-1}

\usepackage{graphicx}
\usepackage{subfigure}
\usepackage{float}

\usepackage{dcolumn}
\usepackage{bm}
\usepackage{amssymb}
\usepackage{latexsym}
\usepackage{booktabs}
\usepackage{amsmath}
\usepackage{multirow}
\usepackage{url}
\usepackage{changes}
\usepackage{soul} 
\usepackage{color, xcolor}
\usepackage[colorlinks=true, linkcolor=red, citecolor=blue]{hyperref}

\usepackage[normalem]{ulem}
\usepackage{array}
\usepackage{enumerate}

\def\be{\begin{equation}}
\def\ee{\end{equation}}

\def\bea{\begin{eqnarray}}
\def\eea{\end{eqnarray}}
\newcommand{\bes}{\begin{equation*}}
\newcommand{\ees}{\end{equation*}}
\newcommand{\beqa}{\begin{eqnarray}}
\newcommand{\eeqa}{\end{eqnarray}}
\newcommand{\ksm}{~{\rm km}~{\rm s}^{-1}~{\rm Mpc}^{-1}}
\newcommand{\rmd}{{\rm d}}

\begin{document}


\title{Gravitational wave standard sirens from GWTC-3 combined with DESI DR2 and DESY5: A late-universe probe of the Hubble constant and dark energy}

\author{Ji-Yu Song}
\affiliation{Liaoning Key Laboratory of Cosmology and Astrophysics, College of Sciences, Northeastern University, Shenyang 110819, China}
\author{Guo-Hong Du}
\affiliation{Liaoning Key Laboratory of Cosmology and Astrophysics, College of Sciences, Northeastern University, Shenyang 110819, China}
\author{Tian-Nuo Li}
\affiliation{Liaoning Key Laboratory of Cosmology and Astrophysics, College of Sciences, Northeastern University, Shenyang 110819, China}
\author{Ling-Feng Wang}
\affiliation{School of Physics and Optoelectronic Engineering, Hainan University, Haikou 570228, China}
\author{Jing-Zhao Qi}
\affiliation{Liaoning Key Laboratory of Cosmology and Astrophysics, College of Sciences, Northeastern University, Shenyang 110819, China}
\author{Jing-Fei Zhang}
\affiliation{Liaoning Key Laboratory of Cosmology and Astrophysics, College of Sciences, Northeastern University, Shenyang 110819, China}
\author{Xin Zhang}\thanks{Corresponding author: \href{mailto:zhangxin@neu.edu.cn}{zhangxin@neu.edu.cn}}
\affiliation{Liaoning Key Laboratory of Cosmology and Astrophysics, College of Sciences, Northeastern University, Shenyang 110819, China}
\affiliation{MOE Key Laboratory of Data Analytics and Optimization for Smart Industry, Northeastern University, Shenyang 110819, China}
\affiliation{National Frontiers Science Center for Industrial Intelligence and Systems Optimization, Northeastern University, Shenyang 110819, China}

\begin{abstract}

Recently, the combination of the Dark Energy Spectroscopic Instrument (DESI) Data Release 2 (DR2) baryon acoustic oscillation (BAO) data and the Planck cosmic microwave background (CMB) measurements has shown a $\sim3\sigma$ preference for a dynamical dark energy model with a phantom-crossing behavior. However, such a phantom-crossing dark energy evolution from $w<-1$ to $w>-1$ further exacerbates the already severe Hubble tension in the $\Lambda$CDM model. Moreover, there exists a $\sim2\sigma$ tension between the DESI DR2 BAO and CMB datasets. Therefore, it is essential to measure the Hubble constant and dark-energy equation-of-state (EoS) parameters using only late-universe observations. In this work, we investigate a novel late-universe data combination: gravitational-wave (GW) standard sirens, BAO, and Type Ia supernovae (SNe Ia). This combination provides a fully distance-ladder- and CMB-independent determination of the Hubble constant and the dark-energy EoS. Using 47 GW standard sirens from the third Gravitational-Wave Transient Catalog, the DESI DR2 BAO data, and DESY5 SNe Ia data, in the $w_0w_a$CDM model, we obtain $H_0=74.8^{+6.3}_{-8.9}\ {\rm km\ s^{-1}\ Mpc^{-1}}$, $\Omega_{\rm m}=0.320^{+0.015}_{-0.012}$, $w_0=-0.775^{+0.072}_{-0.074}$, and $w_a=-0.80\pm0.47$, indicating a mild phantom-crossing behavior within the $1\sigma$ credible interval with an $H_0$ value consistent with the distance ladder measurements. Our analysis demonstrates the power of GW standard sirens in breaking parameter degeneracies, and for the first time, achieves robust joint constraints on the Hubble constant and the dark-energy EoS parameters based on real GW data.

\end{abstract}


\maketitle

\section{Introduction}\label{sec:Introduction}

Over the past two decades, the $\Lambda$CDM model has gained widespread acceptance as a standard cosmological model due to its success in explaining a broad range of cosmological observations and its stunning precision when fitting the cosmic microwave background (CMB) observations. However, some severe tensions between the early- and the late-universe observations have arisen in the $\Lambda$CDM model, such as the well-known Hubble tension. The global fit of the CMB data in the $\Lambda$CDM model yields $H_0 = 67.66 \pm 0.42\ \ksm$~\cite{Planck:2018vyg}, while the distance-ladder determination based on JWST and HST observations gives $H_0 = 73.18 \pm 0.88\ \ksm$, corresponding to a $\sim6\sigma$ tension between these two measurements~\cite{Riess:2025chq}. Numerous studies have examined possible systematics in the observations \cite{Planck:2016tof,Jones:2018vbn,Kenworthy:2019qwq,Riess:2019cxk,Riess:2021jrx,Castello:2021uad,Planck:2019evm,Riess:2024vfa,Li:2024pjo,Pascale:2024qjr,Li:2024yoe,Riess:2024ohe}, but none can explain this large discrepancy. Likewise, no extended cosmological model can simultaneously resolve the Hubble tension and agree well with observations \cite{Guo:2018ans,DAmico:2020ods,Hill:2020osr}.

Recently, the Dark Energy Spectroscopic Instrument (DESI) \cite{DESI:2016fyo} has been gradually releasing data, sparking wide discussions \cite{Wu:2024faw,Li:2024qso,Du:2024pai,Li:2024qus,Ye:2025ark,Pang:2025lvh,Wu:2025wyk,Li:2025owk,Du:2025iow,Feng:2025mlo,Ling:2025lmw,Li:2025eqh,Li:2025dwz,Li:2025htp,Du:2025xes,Wu:2025vfs,Zhou:2025nkb,Zhang:2025dwu,Li:2025muv,Li:2025vqt,Li:2025ops,Halder:2025ytq,Pan:2025qwy,Yang:2025uyv,Cai:2025mas,Huang:2025som,Wang:2024dka,Pedrotti:2025ccw,Jiang:2024viw,Jiang:2024xnu,Giare:2024smz,Liu:2024yib,Colgain:2025nzf,Yao:2025kuz,Giare:2025ath,Yao:2025twv,Liu:2025myr,Li:2025vuh,Cheng:2025yue}. The combination of DESI Data Release 1 (DR1) baryon acoustic oscillation (BAO) data with the CMB data reveals a $2.6\sigma$ deviation from the $\Lambda$CDM model, preferring dynamical dark energy models with phantom-crossing behavior \cite{DESI:2024mwx}. This deviation increases to $3.1\sigma$ with the DESI Data Release 2 (DR2) \cite{DESI:2025zgx}. Adding Type Ia supernovae (SNe Ia) data sustains or enhances this deviation, reaching $2.8$--$4.2\sigma$ depending on the SNe Ia sample used \cite{DESI:2025zgx}. Later, Ref.~\cite{Ye:2025ark} identified a $\sim 2\sigma$ discrepancy between the DESI BAO and CMB data in the $\Lambda$CDM model using both the suspiciousness and goodness-of-fit tension metrics and concluded that three fundamental assumptions in the $\Lambda$CDM model need to be modified to resolve this tension, representing the potential entry points for new physics. However, investigations into new physics, in light of a $\sim2\sigma$ discrepancy between early- and late-universe data, require careful consideration. Particularly, dynamical dark energy with phantom-crossing behavior from $w<-1$ to $w>-1$ could exacerbate the already significant Hubble tension in the $\Lambda$CDM model. Moreover, when exploring extended models of the $\Lambda$CDM model, incorporating the Hubble constant ($H_0$) measurement results from the distance ladder into the CMB+BAO+SNe Ia data will also influence the preference for the model \cite{Pang:2025lvh}. Therefore, jointly constraining $H_0$ and the dark energy equation-of-state (EoS) parameters using only late-universe observations, independent of the distance ladder, is of great significance, see Refs.~\cite{Wu:2025wyk, Liu:2025mub, Nagpal:2025omq} for recent discussions.

Traditional late-universe probes such as BAO and SNe Ia provide high-precision measurements and abundant observational data. However, they can only measure relative distances and therefore require external calibration, such as from CMB or the distance ladder, to determine $H_0$. In recent years, several emerging probes have enabled measurements of $H_0$ independent of CMB and distance ladder, including strong gravitational lensing time delays \cite{H0LiCOW:2019xdh, Birrer:2020tax, Wang:2021kxc, TDCOSMO:2025dmr, Liu:2025ayr}, cosmic chronometers \cite{Jimenez:2001gg, Cai:2021weh, Cai:2022dkh}, quasars \cite{Lusso:2020pdb, Wei:2020suh}, gamma-ray bursts \cite{Demianski:2016zxi, Amati:2018tso, Demianski:2019vzl, Du:2025csv}, fast radio bursts \cite{Zhao:2020ole,Wu:2021jyk,Zhao:2022bpd,Zhao:2022yiv,Zhang:2023pqs,Zhang:2023gye,Sun:2024huw,Zhang:2024rra,Wang:2025ugc,Gao:2025fcr,Sun:2025cdq}, and gravitational-wave (GW) standard sirens \cite{LIGOScientific:2017adf, LIGOScientific:2017vwq, LIGOScientific:2021aug, LIGOScientific:2025jau}. Compared with emerging probes based on the electromagnetic (EM) observation, GW standard sirens have the advantage of directly measuring luminosity distances without any calibration, relying solely on general relativity, which is widely believed to be a promising cosmological probe for resolving the Hubble tension \cite{Schutz:1986gp,
Zhao:2010sz,Cai:2016sby,Chen:2017rfc,Wang:2018lun,Zhang:2018byx,Feeney:2018mkj,Zhang:2019ylr,Zhang:2019loq,Wang:2019tto,Gray:2019ksv,Yu:2020vyy,Jin:2020hmc,Zhu:2021aat,Zhu:2021bpp,Qi:2021iic,Jin:2021pcv,Wang:2021srv,LIGOScientific:2021aug,Mastrogiovanni:2021wsd,Wu:2022dgy,Song:2022siz,Jin:2022tdf,Jin:2022qnj,Wang:2022oou,Jin:2023tou,Yu:2023ico,Jin:2023sfc,Han:2023exn,Jin:2023zhi,Li:2023gtu,Yun:2023ygz,Mastrogiovanni:2023emh,Mastrogiovanni:2023zbw,Feng:2024mfx,Feng:2024lzh,Han:2024sxm,Dong:2024bvw,Xiao:2024nmi,Zhu:2024qpp,Zheng:2024mbo,Giare:2024syw,Han:2025fii,Feng:2025wbz,Song:2025ddm,Xiao:2025mcg,Dong:2025ikq,Zhang:2025yhi,Du:2025odq,Zhan:2025jqg,Li:2025eig,Su:2025zuc}; see Ref.~\cite{Jin:2025dvf} for an up-to-date review of using GW standard sirens to constrain cosmological parameters. 

Currently, the LIGO-Virgo-KAGRA (LVK) O1--O4a runs have detected more than 300 GW events, among which 142 events have been used to constrain $H_0$ with a precision of about 14\% \cite{LIGOScientific:2025jau}. However, due to the limited measurement precisions and the relatively low redshifts of current GW data, they are not yet able to place meaningful constraints on cosmological parameters other than $H_0$. A work~\cite{Pierra:2025hoc} released concurrently with this article reconstructs the Hubble parameter based on 137 GW events and investigates the redshift ranges where the current GW data can provide strong constraints on the expansion history of the universe. In this work, we note that the absolute luminosity-distance measurements from GW standard sirens are complementary to the relative distance measurements from BAO and SNe Ia. Particularly, the independent measurement of $H_0$ from GW standard sirens can break the degeneracy between $H_0$ and the sound horizon $r_{\rm d}$ in BAO analyses, as well as the degeneracy between $H_0$ and the absolute magnitude $M_B$ in SNe Ia. Therefore, by combining GW standard sirens with BAO and SNe Ia, we achieve, for the first time, a robust joint constraint on the Hubble constant and the dark-energy EoS parameters based on real-observed GW standard siren data.

In particular, we explore the cosmological parameter constraint capability of combining 47 GW standard siren events from the third Gravitational-Wave Transient Catalog (GWTC-3) \cite{Poggiani:2025pyy}, the BAO data from DESI DR2, and SNe Ia data from Dark Energy Survey Year 5 (DESY5) \cite{DES:2024upw, DES:2024jxu} in the $\Lambda$CDM and $w_0w_a$CDM models. We use the EM counterpart of GW170817 to obtain its redshift, and combine the GLADE+ catalog \cite{Dalya:2021ewn} and GW population models to infer the redshifts of the remaining 46 GW events without EM counterparts. We treat $r_{\rm d}$ and $M_B$ as free parameters to avoid the reliance on CMB and the distance ladder.  

This paper is organized as follows. In Sect.~\ref{sec:method}, we outline the methods and data used in this paper. In Sect.~\ref{sec:results}, we present the results and make relevant discussions. Finally, in Sect.~\ref{sec:conclusions}, we summarize the key findings and conclude the study.

\section{Methodology and data}\label{sec:method}

In this section, we briefly introduce the cosmological models adopted in this work, followed by a description of the observational datasets and the corresponding likelihood functions used in our analysis. Finally, we present the Markov Chain Monte Carlo (MCMC) sampling method employed for parameter estimation, together with the model assumptions and prior settings.

\subsection{Distances and cosmological models}

In this work, we consider two representative cosmological models: the $\Lambda$CDM model and the $w_0w_a$CDM model \cite{Chevallier:2000qy, Linder:2002et}.

In the Friedmann-Lema\^{\i}tre-Robertson-Walker metric, the comoving distance is defined as
\begin{equation}\label{DC}
D_{\rm C}(z)=\int_{0}^{z} \frac{c{\rm d}z'}{H(z')},
\end{equation}
where $H(z)$ is the Hubble parameter and $c$ is the speed of light. The transverse comoving distance is given by \cite{Hogg:1999ad}
\begin{equation}\label{eq:DM}
D_{\rm M}(z)=\left\{
\begin{aligned}
&\frac{c}{H_0\sqrt{\Omega_K}}\sinh{\left[\frac{H_0\sqrt{\Omega_K}}{c}D_{\rm C}(z)\right]} & {\rm if}\ \Omega_{K}>0 ,  \\
&D_{\rm C}(z)                                                                             & {\rm if}\ \Omega_{K}=0 , \\
&\frac{c}{H_0\sqrt{-\Omega_K}}\sin{\left[\frac{H_0\sqrt{-\Omega_K}}{c}D_{\rm C}(z)\right]}  & {\rm if}\ \Omega_{K}<0 ,
\end{aligned}
\right.
\end{equation}
where $\Omega_K$ is the spatial curvature parameter. The luminosity distance and Hubble distance are given by
\begin{align}\label{eq:DLandDH}
D_{\rm L}(z)=D_{\rm M}(z)(1+z);\  D_{\rm H}(z)=c/H(z).
\end{align}

In the $\Lambda$CDM model, the dark-energy EoS is a constant, $w=-1$, and $H(z)$ is given by
\begin{equation}
    H(z)=H_0\sqrt{\Omega_{\rm m}(1+z)^3+(1-\Omega_{\rm m})},
\end{equation}
where $\Omega_{\rm m}$ is the density parameter of matter.

In the $w_0w_a$CDM model, the dark-energy EoS can deviate from $w=-1$ and is parameterized as $w(z)=w_0+w_az/(1+z)$, and $H(z)$ is written as
\begin{equation}
\begin{aligned}
    H(z)=&H_0\bigg[\Omega_{\rm m}(1+z)^3\\&+(1-\Omega_{\rm m})(1+z)^{3(1+w_0+w_a)}\exp\bigg(-\frac{3w_az}{1+z}\bigg)\bigg]^{1/2}.
\end{aligned}
\end{equation}

\subsection{Gravitational-wave standard sirens}

In this work, we use the same 47 GW events as adopted in Ref.~\cite{LIGOScientific:2021aug}. The selected GW sample includes GW events in the LVK O1--O3 runs with a network signal-to-noise ratio (SNR) greater than 11 and a false-alarm rate (FAR) below 0.25, consisting of 42 binary black hole (BBH) mergers, 3 neutron star-black hole (NSBH) mergers, and 2 binary neutron star (BNS) mergers. All GW data are publicly available from the Gravitational Wave Open Science Center\footnote{\url{https://gwosc.org/eventapi/html/allevents/}}. We did not use the latest O4a GW data because the GW detector, Virgo, was not working during this run, and all GW events were detected by two LIGOs, resulting in relatively weak spatial localization of the GW events. For the latest constraints on the cosmic expansion rate and modified GW propagation using 142 GW events from the LVK O1--O4a runs, refer to Ref.~\cite{LIGOScientific:2025jau}.

Obtaining redshifts of GW events is crucial for using them in cosmological model constraints. In this work, we consider both ``bright siren'' and ``dark siren'' methods to obtain redshifts of GW sources. The bright siren method relies on the detection of EM counterparts, which help identify host galaxies of GW sources and enable precise spectroscopic measurement of their redshifts. Since EM observations are independent of GW observations, the likelihood can be written as
\begin{equation}
    \mathcal{L}
    =\mathcal{L}_1(x_{\rm EM}|z,\Omega)\mathcal{L}_2(x_{\rm GW}|D_{\rm L},m_1,m_2,\alpha,\delta),
\end{equation}
where $\mathcal{L}_1(x_{\rm EM}|z,\Omega)$ is the likelihood of EM observations, quantifying the measurement uncertainties of the redshift $z$ and the sky location $\Omega$ of the GW event. Similarly, the likelihood of GW detections, $\mathcal{L}_2(x_{\rm GW}|D_{\rm L},m_1,m_2,\alpha,\delta)$, characterizes the measurement uncertainties of the luminosity distance $D_{\rm L}$, the primary mass $m_1$, the secondary mass $m_2$, the right ascension $\alpha$, and the declination $\delta$ of the GW event. Among the GW events considered in this paper, GW170817 is the only bright siren, and we obtain its redshifts by spectroscopic observation of its host galaxy NGC 4993 \cite{LIGOScientific:2017adf, LIGOScientific:2017vwq, LIGOScientific:2017zic}. For the remaining 46 GW events, we infer their redshifts by the dark siren method.

For GW events without EM counterparts, their sky positions rely solely on GW observations and are therefore poorly localized; even in the best cases, the uncertainties remain as large as $\sim 20\ \mathrm{deg}^2$, preventing unique identification of their host galaxies. Following Refs.~\cite{Gray:2019ksv, LIGOScientific:2021aug}, we employ the GLADE+ galaxy catalog to construct, for each GW event, a redshift prior based on the galaxy distribution within its sky localization region. In addition, we incorporate an additional redshift prior based on population models that describe the distribution of source-frame masses and redshifts of GW events \cite{Taylor:2011fs, Mastrogiovanni:2021wsd, Ezquiaga:2022zkx, Gray:2023wgj, Mastrogiovanni:2023emh}. Such source-frame mass distribution models can help break the degeneracies between masses and redshifts inherent in GW observations, and these effects become more significant when the mass distribution exhibits distinct peaks or other sharp features \cite{Ezquiaga:2022zkx}. In summary, we combine the redshift priors derived from the GLADE+ catalog and the GW source population models to construct a joint redshift prior for each dark siren event, and use it within a hierarchical Bayesian framework to perform a joint inference of both the population-model parameters and the cosmological parameters. The likelihood of dark sirens is given by
\begin{equation}\label{eq:GWlikelihood}
    \begin{aligned}
        \mathcal{L}(\{x_{\rm GW}\}|\boldsymbol{\Lambda})\propto\prod_{i}^{N_{\rm GW}}\frac{\int{\rm d}\boldsymbol{\theta}\mathcal{L}(x_{{\rm GW},i}|\boldsymbol{\theta},\boldsymbol{\Lambda})\frac{{\rm d}N_{\rm GW}}{{\rm d}\boldsymbol{\theta}{\rm d}t}(\boldsymbol{\Lambda})}{\int{\rm d}\boldsymbol{\theta}p_{\rm det}(\boldsymbol{\theta},\boldsymbol{\Lambda})\frac{{\rm d}N_{\rm GW}}{{\rm d}\boldsymbol{\theta}{\rm d}t}(\boldsymbol{\Lambda})},
    \end{aligned}
\end{equation}
where $N_{\rm GW}$ is the number of GW events in the GW dataset $\{x_{\rm GW}\}$. $\boldsymbol{\theta}$ represent GW source parameters in the detector frame, including $D_{\rm L}$, $m_1$, $m_2$, $\alpha$, and $\delta$. $\boldsymbol{\Lambda}$ are hyperparameters, including cosmological parameters and population-model parameters. $p_{\rm det}(\boldsymbol{\theta},\boldsymbol{\Lambda})$ represent the detection probability of GW events. ${\rm d}N_{\rm GW}/{\rm d}\boldsymbol{\theta}{\rm d}t(\boldsymbol{\Lambda})$ is the GW event rate in the detector frame, which is related to the source-frame GW event rate ${\rm d}N_{\rm GW}/{\rm d}\boldsymbol{\theta}_{\rm s}{\rm d}t_{\rm s}(\boldsymbol{\Lambda})$ via
\begin{equation}
    \frac{{\rm d}N}{{\rm d}\boldsymbol{\theta}{\rm d}t}=\frac{{\rm d}N}{{\rm d}\boldsymbol{\theta}_{\rm s}{\rm d}t_{\rm s}}\frac{1}{1+z}\frac{1}{\frac{\partial D_{\rm L}}{\partial z}(1+z)^2},
\end{equation}
where $1/(1+z)$ comes from the difference between source-frame and detector-frame time, and $\partial D_{\rm L}/\partial z(1+z)^2$ is the Jacobian for the change of variables between the detector and the source frames. GW event rate in the source frame can be written as
\begin{equation}\label{eq:eventrate}
\begin{aligned}
    \frac{\rmd N_{\rm GW}}{\rmd \boldsymbol{\theta}_{\rm s}\rmd z\rmd t_{\rm s}}&=\frac{\rmd N_{\rm GW}}{\rmd m_{\rm s,1}\rmd m_{\rm s,2}\rmd \Omega\rmd z\rmd t_{\rm s}}\\
    &=\int_{M_{\rm min}}^{M_{\rm max}}\frac{\rmd N_{\rm GW}}{\rmd m_{\rm s,1}\rmd m_{\rm s,2}\rmd \Omega\rmd z\rmd t_{\rm s}\rmd M}\rmd M\\
    &=\int_{M_{\rm min}}^{M_{\rm max}}\rmd M \frac{\rmd N_{\rm GW}}{\rmd N_{\rm gal}\rmd m_{s,1}\rmd m_{s,2}\rmd t_{\rm s}}\frac{\rmd N_{\rm gal}}{\rmd z\rmd \Omega\rmd M},
\end{aligned}
\end{equation}
where $M$ is the absolute magnitude of galaxies, and $\rmd N_{\rm GW}/(\rmd N_{\rm gal}\rmd m_{s,1}\rmd m_{s,2}\rmd t_{\rm s})$ denotes the GW event rate per galaxy, which can be expressed as
\begin{equation}
    \frac{\rmd N_{\rm GW}}{\rmd N_{\rm gal}\rmd m_{s,1}\rmd m_{s,2}\rmd t_{\rm s}}\propto \psi(z|\boldsymbol{\Lambda}_{z})p_{\rm pop}(m_{\rm s,1},m_{\rm s,2}|\boldsymbol{\Lambda}_{m})\bigg(\frac{L}{L_{*}}\bigg)^{\epsilon},
\end{equation}
where $\psi(z|\boldsymbol{\Lambda}_{z})$ is the redshift evolution function of GW events, describing how the probability of a galaxy hosting a GW source varies with redshift. The term $p_{\rm pop}(m_{\rm s,1},m_{\rm s,2}|\boldsymbol{\Lambda}_{m})$ characterizes the mass distribution of GW sources. The luminosity weight factor $(L/L{})^{\epsilon}$ accounts for the correlation between galaxy luminosity and the likelihood of hosting GW events, with $L_{*}$ being the characteristic luminosity of the Schechter function. A positive value $\epsilon>0$ implies that more luminous galaxies are more likely to host GW events, while $\epsilon=0$ corresponds to the case with no luminosity dependence. In this work, we fix $\epsilon=1$. According to Refs.~\cite{LIGOScientific:2021aug, LIGOScientific:2025jau}, $\epsilon$ shall not influence the cosmological parameters inference largely.

The term $\rmd N_{\rm gal}/(\rmd z\rmd \Omega\rmd M)$ denotes the galaxy number density. Owing to the flux limits of survey telescopes, galaxy catalogs are typically complete under an apparent magnitude threshold $m_{\rm th}$. To reconstruct the true galaxy number density in the universe, one must account for the galaxies that fall out of this threshold. Accordingly, the galaxy number density can be written as
\begin{equation}\label{eq:completeness}
    \frac{\rmd N_{\rm gal}}{\rmd z\rmd \Omega\rmd M}=\frac{\rmd N_{\rm gal, cat}}{\rmd z\rmd \Omega\rmd M}+\frac{\rmd N_{\rm gal, out}}{\rmd z\rmd \Omega\rmd M},
\end{equation}
where $\rmd N_{\rm gal, out}/(\rmd z,\rmd \Omega,\rmd M)$ is the completeness correction term, given by
\begin{equation}
    \begin{aligned}
        \frac{\rmd N_{\rm gal, out}}{\rmd z\rmd \Omega\rmd M}\propto&{\rm Sch}(M|\alpha,\phi_*,M_*)\\&\times\frac{\rmd V_{\rm c}}{\rmd\Omega\rmd z}\Theta[M>M_{\rm thr}(z,m_{\rm th},H_0)],
    \end{aligned}
\end{equation}
where ${\rm Sch}(M|\alpha,\phi_*, M_*)$ is the Schechter function that describes the distribution of galaxy absolute magnitudes. In this work, we use the $K$ band galaxies of GLADE+, and assume that the $K$-band absolute magnitude of galaxies is well described by the Schechter function with parameters provided in Ref.~\cite{Kochanek:2000im}. The term $\rmd V_{\rm c}/(\rmd \Omega,\rmd z)$ corresponds to the assumption that galaxies are uniformly distributed in the comoving volume. The step function $\Theta[M > M_{\rm th}(z,m_{\rm th},H_0)]$ equals unity when $M > M_{\rm th}$ and vanishes otherwise.

The term $\rmd N_{\rm gal, cat}/(\rmd z\rmd \Omega\rmd M)$ represents the catalog term, given by
\begin{equation}\label{eq:catalogterm}
\begin{aligned}
    \frac{\rmd N_{\rm gal, cat}}{\rmd z\rmd \Omega\rmd M}\propto&\sum_{j}^{N_{\rm gal}(\Omega)}\mathcal{N}(z_j,\sigma_j)\frac{\rmd V_{\rm c}}{\rmd z}\\&\times\delta[M-M(m_j,H_0,z_j)]\delta(\Omega-\Omega_j),
\end{aligned}
\end{equation}
where $\mathcal{N}$ is the Gaussian function, and $N_{\rm gal}(\Omega)$ is the galaxy number in a given sky direction. In this work, we use the public pipeline \texttt{ICAROGW} to conduct dark siren analysis \cite{Mastrogiovanni:2023zbw}, which enhances the calculation effectiveness by using the Monte Carlo integration method and preconstructs an interpolated function of the galaxy redshift distribution on each pixel of the catalog. The catalog is pixelized using \texttt{healpy} and \texttt{HEALPix} package \cite{Gorski:2004by, Zonca:2019vzt} with a resolution parameter of $n_{\rm side}=64$.

\subsection{Baryon acoustic oscillations}

In this work, we use the BAO measurements from DESI DR2, which are based on observations of galaxies, quasars, and the Lyman-$\alpha$ forest across seven redshift bins \cite{DESI:2025zgx}. Compared with DESI DR1 \cite{DESI:2024mwx}, the DESI DR2 BAO measurements based on a longer observational time and a significantly larger number of tracers. Specifically, DR2 data covers nearly three years of observations (one year for DR1) and includes more than 14 million spectroscopically confirmed objects, roughly doubling the sample size of DR1.

The BAO measurements provide distance information relative to the sound horizon $r_{\rm d}$, including the transverse comoving distance $D_{\rm M}(z)/r_{\rm d}$, the Hubble distance $D_{\rm H}(z)/r_{\rm d}$, or their combination $D_{\rm V}(z)/r_{\rm d}\equiv\big(zD_{\rm M}(z)^2D_{\rm H}(z)\big)^{1/3}/r_{\rm d}$. Table~\ref{tab:BAOdata} summarizes the BAO data used in this work. The sound horizon $r_{\mathrm{d}}$ is the comoving sound horizon at the baryon drag epoch, representing the maximum distance that sound waves could propagate before the decoupling of baryons and photons. In this analysis, we treat $r_{\mathrm{d}}$ as a free parameter jointly constrained with the cosmological parameters, to avoid dependence on early-universe observations.

\begin{table}
\renewcommand\arraystretch{1}
\caption{The BAO measurements from DESI. Here, $r_{\rm d}$ is the sound horizon at the drag epoch, $D_{\rm M}$ is the transverse comoving distance, $D_{\rm H}$ is the Hubble distance, and $D_{\rm V}\equiv(zD_{\rm M}^2D_{\rm H})^{1/3}$.}
\label{tab:BAOdata}
\centering
\begin{tabular}{ccc}
\hline\hline
\makebox[0.15\textwidth][c]{Redshift $z$} & \makebox[0.15\textwidth][c]{Quantity} & \makebox[0.15\textwidth][c]{Value}\\
\hline
0.295         & $D_{\rm V}/r_{\rm d}$         & $7.942\pm0.075$            \\
0.510         & $D_{\rm M}/r_{\rm d}$         & $13.588\pm0.167$            \\
0.510         & $D_{\rm H}/r_{\rm d}$         & $21.863\pm0.425$           \\
0.706         & $D_{\rm M}/r_{\rm d}$         & $17.351\pm0.177$            \\
0.706         & $D_{\rm H}/r_{\rm d}$         & $19.455\pm0.330$            \\
0.934         & $D_{\rm M}/r_{\rm d}$         & $21.576\pm0.152$            \\
0.934         & $D_{\rm H}/r_{\rm d}$         & $17.641\pm0.193$           \\
1.321         & $D_{\rm M}/r_{\rm d}$         & $27.601\pm0.318$             \\
1.321         & $D_{\rm H}/r_{\rm d}$         & $14.176\pm0.221$          \\
1.484         & $D_{\rm M}/r_{\rm d}$         & $30.512\pm0.760$           \\
1.484         & $D_{\rm H}/r_{\rm d}$         & $12.817\pm0.516$           \\
2.330         & $D_{\rm H}/r_{\rm d}$         & $8.632\pm0.101$            \\
2.330         & $D_{\rm M}/r_{\rm d}$         & $38.988\pm0.531$          \\
\hline
\hline
\end{tabular}
\end{table}

The BAO data have been processed such that their likelihood functions can be expressed in a simplified form:
\begin{equation}\label{eq:likelihoodBAOSNeIa}
    \mathcal{L}\propto\exp\bigg[-\frac{1}{2}(\boldsymbol{X}_{\rm the}-\boldsymbol{X}_{\rm obs})^{T}\boldsymbol{C}^{-1}(\boldsymbol{X}_{\rm the}-\boldsymbol{X}_{\rm obs})\bigg],
\end{equation}
where $\boldsymbol{C}$ is the covariance matrix of the data, and $\boldsymbol{X}_{\rm the}$ and $\boldsymbol{X}_{\rm obs}$ is the theory value and observed value of the data. When using the BAO data, $X$ represents $D_{\rm M}(z)/r_{\rm d}$, $D_{\rm H}(z)/r_{\rm d}$, or $D_{\rm V}(z)/r_{\rm d}$. The data and and likelihood function of BAO can be found on the website\footnote{\url{https://github.com/CobayaSampler/bao_data/tree/master}}.

\subsection{Type Ia supernovae}

In this work, we use the SNe Ia sample from DESY5 \cite{DES:2024upw, DES:2024jxu}, which consists of 1829 SNe Ia covering a redshift range from 0.025 to 1.3. As standard candles, SNe Ia can provide measurements of the luminosity distance. Specifically, the SNe Ia data directly offer apparent magnitude measurements in the rest frame $m_B$, which are related to the absolute magnitude $M_B$ and the luminosity distance $D_{\rm L}$ through the following relation:
\begin{equation}
    m_B(z) = M_B +5{\rm log}_{10}(D_{\rm L}(z)/{\rm Mpc})+25.
\end{equation}

Since the absolute magnitude of SNe Ia can be regarded as a constant, the apparent magnitude can thus be used for distance estimation. We treat $M_B$ as a free parameter jointly constrained with the cosmological parameters to avoid dependence on the distance ladder. The likelihood function of the SNe Ia data is similar to that of BAO, as given by Eq.~\eqref{eq:likelihoodBAOSNeIa}, where $X$ denotes the apparent magnitude. The data and likelihood of DESY5 SNe Ia can be found on the website\footnote{\url{https://github.com/des-science/DES-SN5YR}}.

\subsection{Cosmological inference and prior specification}

In this work, we perform MCMC sampling using the Bayesian inference library \texttt{Bilby} \citep{Ashton:2018jfp}, with the \texttt{emcee} sampler \citep{Foreman-Mackey:2012any}. The likelihood function for cosmological parameter inference is the combination of various data likelihoods. For instance, when combining GW standard siren data, BAO data, and SNe Ia data, the logogram of the total likelihood can be expressed as
\begin{equation}
    {\rm log}(\mathcal{L}_{\rm tot}) = {\rm log}(\mathcal{L}_{\rm GW})+{\rm log}(\mathcal{L}_{\rm BAO})+{\rm log}(\mathcal{L}_{\rm SNe}),
\end{equation}
where $\mathcal{L}_{\mathrm{GW}}$, $\mathcal{L}_{\mathrm{BAO}}$, and $\mathcal{L}_{\mathrm{SNe}}$ denote the likelihood functions for the GW standard sirens, BAO, and SNe Ia data, respectively.

When using the GW standard siren data, we jointly constrain the mass-distribution model parameters of BBH events, the redshift-distribution model parameters of all GW events, and the cosmological parameters. The prior ranges of population models are the same as Table~7 of Ref.~\cite{LIGOScientific:2021aug}. We adopt the population model that is most favored by the GWTC-3 GW data \cite{LIGOScientific:2020kqk, LIGOScientific:2021aug}, the Power Law + Peak model, to describe the mass distribution of BBHs. The mass distribution of primary black holes in NSBHs is the same as that of BBHs. We fix the mass distribution model parameters of BNS events and the neutron star part of NSBH events, and the mass distribution of neutron stars in both NSBHs and BNSs is assumed to be uniform between 1 $M_{\odot}$ and 3 $M_{\odot}$. The redshift distribution of all GW events is a phenomenological approach that assumes the binary formation rate follows the star formation rate \cite{Madau:2014bja}. Following Ref.~\cite{LIGOScientific:2021aug}, we neglect the potential evolution of the BBH mass distribution with redshift and the selection effects of the spin distribution. The expected evolution of the BBH mass distribution is below the current statistical uncertainties in the redshift range of GW data \cite{Fishbach:2021mhp, vanSon:2021zpk}, and including the spin distribution does not alter the detection probability by more than a factor of two \cite{Ng:2018neg}. 

We adopt uniform priors for the cosmological parameters as follows: $H_0 \in [20,\ 100]\ \ksm$, $\Omega_{\rm m} \in [0,\ 1]$, $w_0 \in [-3,\ 1]$, and $w_a \in [-3,\ 2]$. When including the SNe Ia data, the absolute magnitude $M_B$ is assigned a uniform prior range of $[-23,\ -17]$, and when including the BAO data, the sound horizon scale $r_{\rm d}$ is set uniformly within $[100,\ 170]\ {\rm Mpc}$.

\section{Results and discussions}\label{sec:results}

\begin{table*}
\centering
\caption{Results of parameter constraints using different datasets in the $\Lambda$CDM and $w_0w_a$CDM models. GW data represent 47 events from GWTC-3 with SNR $>11$ and FAR $<0.25$, including the bright siren, GW170817, and 46 GW events without EM counterparts. We assume the mass distribution of BBHs follows the Power Law + Peak model and jointly infer cosmological and GW population-model parameters, and use GLADE+ to help estimate redshifts of GW events without EM counterparts. BAO data are taken from DESI DR2, and SNe Ia data from DESY5.}
\label{tab:results}
\centering
\renewcommand{\arraystretch}{1.4}
\begin{tabular}{lcccccc}
\hline\hline
\makebox[0.12\textwidth][l]{Data} & \makebox[0.12\textwidth][c]{$H_0\ [\ksm]$} & \makebox[0.12\textwidth][c]{$\Omega_{\rm m}$} & \makebox[0.12\textwidth][c]{$w_0$} & \makebox[0.12\textwidth][c]{$w_a$} & \makebox[0.12\textwidth][c]{$r_{\rm d}\ [{\rm Mpc}]$} & \makebox[0.12\textwidth][c]{$M_B$}\\
\hline
\multicolumn{7}{c}{\textbf{$\Lambda$CDM}} \\
GW & $76.0^{+6.5}_{-11.3}$ & $0.502^{+0.333}_{-0.332}$ & --& -- & -- & --\\
BAO & -- & $0.294\pm0.009$ & --& -- & -- & --\\
SNe Ia & -- & $0.353^{+0.017}_{-0.016}$ & --& -- & -- & --\\
BAO+SNe Ia & -- & $0.310^{+0.008}_{-0.009}$ & --& -- & -- & -- \\
GW+BAO & $73.1^{+5.6}_{-8.7}$ & $0.294\pm0.010$ & --& -- & $140.9^{+15.0}_{-14.2}$ & --\\
GW+SNe Ia & $75.6^{+6.4}_{-10.8}$ & $0.353^{+0.016}_{-0.017}$ & --& --& -- & $-19.2^{+0.2}_{-0.3}$\\
GW+BAO+SNe Ia & $76.1^{+6.1}_{-11.5}$ & $0.310\pm0.008$& --& -- & $134.1^{+17.4}_{-14.7}$ & $-19.2^{+0.2}_{-0.3}$ \\
\hline
\multicolumn{7}{c}{\textbf{$w_0w_a$CDM}} \\
GW & $76.1^{+6.8}_{-12.2}$ & $0.524^{+0.392}_{-0.249}$ & $-0.947^{+1.453}_{-1.274}$ & $-0.50^{+1.71}_{-1.72}$ & -- & --\\
BAO & -- & $0.346^{+0.057}_{-0.015}$ & $-0.466^{+0.381}_{-0.192}$ & $<-1.23$ & -- & --\\
SNe Ia & -- & $0.374^{+0.072}_{-0.024}$ & $-0.821^{+0.143}_{-0.113}$ & $<-1.31$ & -- & --\\
BAO+SNe Ia & -- & $0.319^{+0.016}_{-0.011}$ & $-0.774^{+0.069}_{-0.076}$ & $-0.79^{+0.48}_{-0.48}$ & -- & --\\
GW+BAO & $74.7^{+6.7}_{-8.8}$ & $0.296\pm0.010$ & $-0.934^{+0.092}_{-0.091}$ & $-0.52^{+1.67}_{-1.66}$ & $135.5^{+15.3}_{-15.0}$ & --\\
GW+SNe Ia & $74.1^{+7.0}_{-9.5}$ & $0.363^{+0.079}_{-0.021}$ & $-0.806^{+0.137}_{-0.117}$ & $<-1.10$ & -- & $-19.2^{+0.2}_{-0.3}$ \\
GW+BAO+SNe Ia & $74.8^{+6.3}_{-8.9}$ & $0.320^{+0.015}_{-0.012}$ & $-0.775^{+0.072}_{-0.074}$ & $-0.80\pm0.47$ & $133.3\pm14.4$ & $-19.2\pm0.2$\\
\hline\hline
\end{tabular}
\end{table*}

\begin{figure}
    \centering
    \includegraphics[width=1\linewidth]{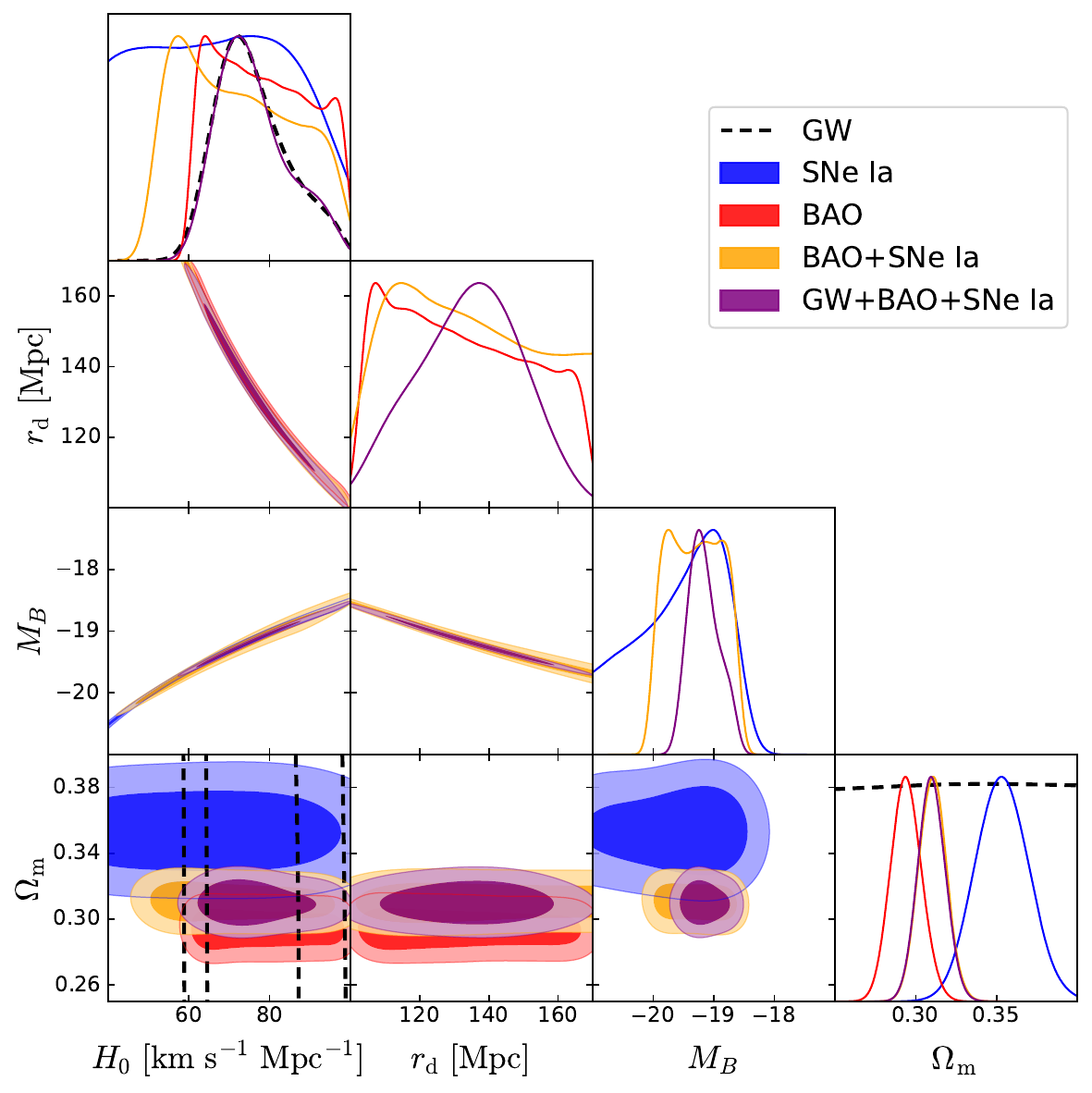}
    \caption{Posterior distributions (68$\%$ and 95$\%$ credible regions) of $H_0$, $r_{\rm d}$, $M_B$, and $\Omega_{\mathrm{m}}$ in the $\Lambda$CDM model, obtained from GW, SNe Ia, BAO, BAO+SNe Ia, and GW+BAO+SNe Ia data.}
    \label{fig:H0rdMB}
\end{figure}

\begin{figure}
    \centering
    \includegraphics[width=1\linewidth]{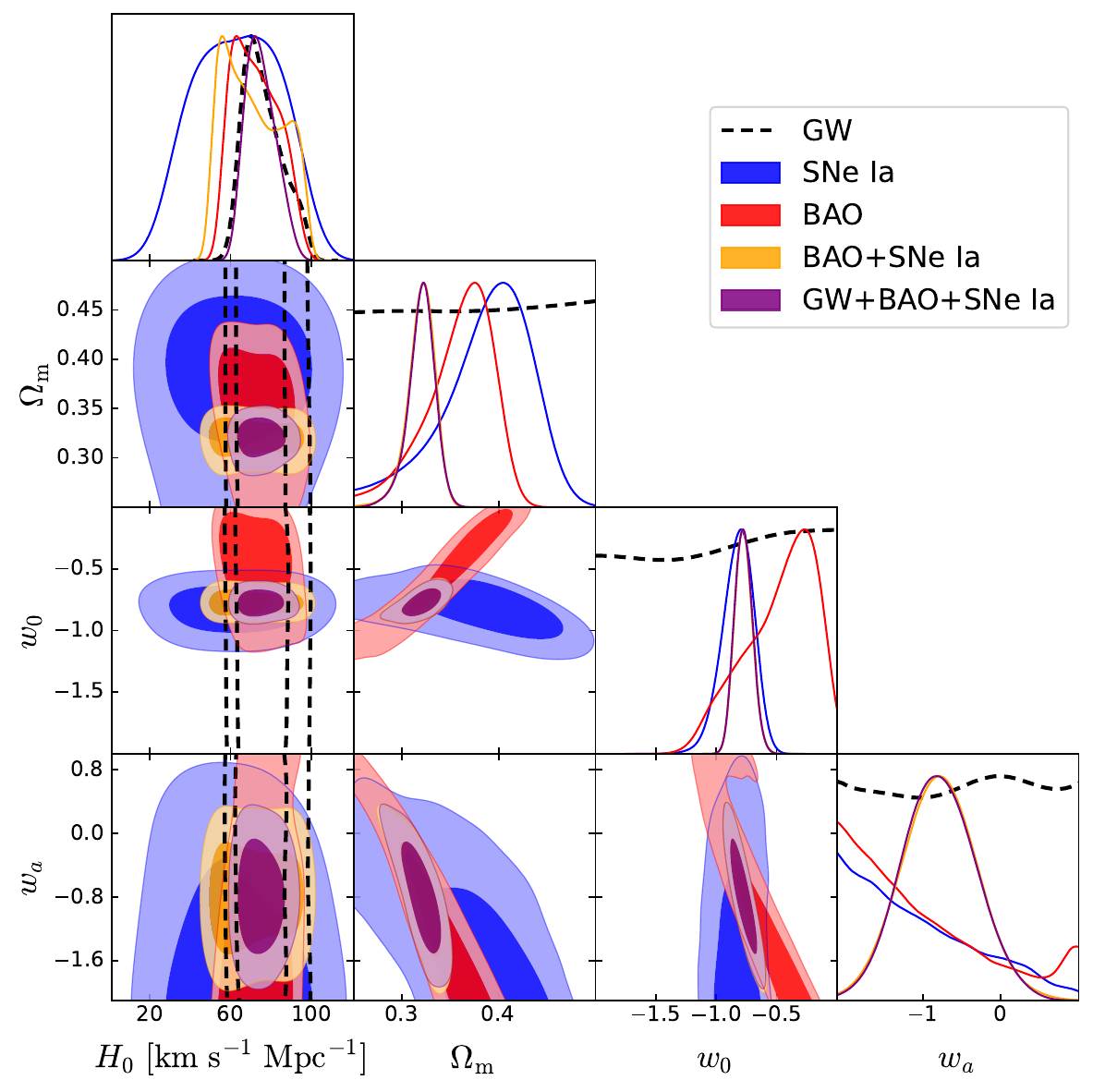}
    \caption{Posterior distributions (68$\%$ and 95$\%$ credible regions) of $H_0$, $\Omega_{\mathrm{m}}$, $w_0$, and $w_a$ in the $w_0w_a$CDM model, obtained from GW, SNe Ia, BAO, BAO+SNe Ia, and GW+BAO+SNe Ia data.}
    \label{fig:H0w0wa}
\end{figure}

\begin{figure}
    \centering
    \includegraphics[width=1\linewidth]{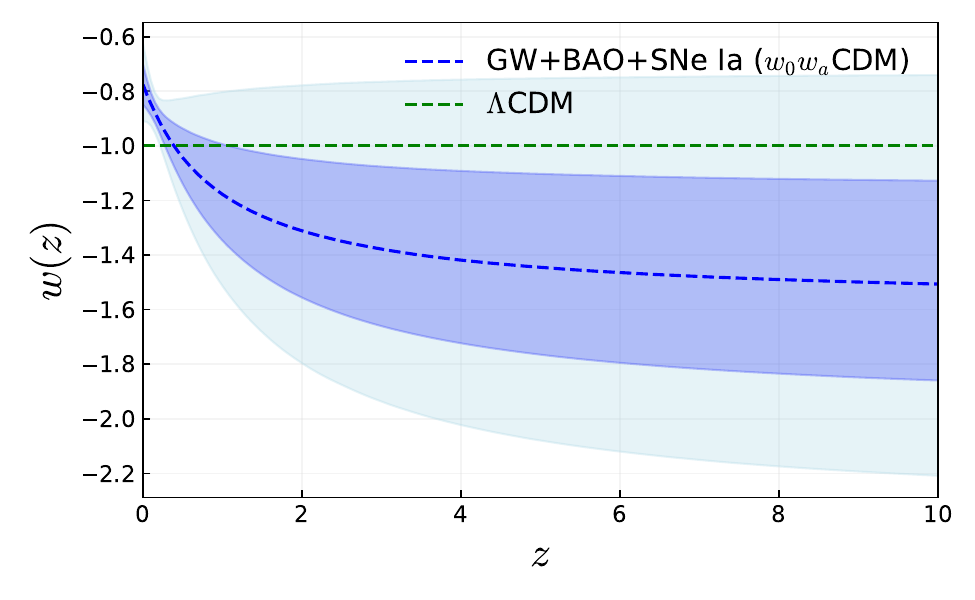}
    \caption{The dark energy EoS across the redshift. The blue lines represent the results of GW+BAO+SNe Ia in the $w_0w_a$CDM model, with the blue and light blue shadow regions indicating the 68$\%$ and 95$\%$ credible regions. The green line represents the situation of the $\Lambda$CDM model with $w=-1$.}
    \label{fig:wz}
\end{figure}

In this section, we present our results and provide relevant discussions. We consider two representative cosmological models to demonstrate the constraining power of the GW standard sirens when combined with BAO and SNe Ia observations. We summarize the results of cosmological parameter constraints in Table~\ref{tab:results}.

We first analyzed the results in the $\Lambda$CDM model. In Fig.~\ref{fig:H0rdMB}, we present the cosmological parameter constraints obtained from the joint analysis of GW, BAO, and SNe Ia data in the $\Lambda$CDM model. This figure clearly demonstrates that the inclusion of GW data can effectively break the degeneracies between $H_0$, $\Omega_{\rm m}$, $M_B$, and $r_{\rm d}$, and the combination of BAO and SNe Ia data can significantly improve the constraint on $\Omega_{\rm m}$. Therefore, using GW+BAO+SNe Ia can achieve a joint well constraint on $H_0$ and $\Omega_{\rm m}$ in the $\Lambda$CDM model, with $H_0=76.1^{+6.1}_{-11.5}\ \ksm$ and $\Omega_{\rm m}=0.310\pm0.008$. We further find that the joint analysis of GW with BAO and SNe Ia data improves the constraints on $H_0$ compared with that of the GW data only, which is primarily due to the strong constraints that BAO and SNe Ia provide on $\Omega_{\rm m}$, which help reduce the parameter degeneracies. Compared with GW data only, GW+BAO, GW+SNe Ia, and GW+BAO+SNe Ia data improve the $H_0$ constraint precisions by 16$\%$, 3$\%$, and 1$\%$, respectively. We noticed that adding both BAO and SNe Ia to the GW data does not improve the $H_0$ constraint precision as much as adding only BAO or SNe Ia data, mainly because adding both BAO and SNe Ia adds two additional free parameters, which may reduce the constraint on $H_0$.

We then proceed to analyze the results in the $w_0w_a$CDM model. In Fig.~\ref{fig:H0w0wa}, we show the posteriors of cosmological parameters in the $w_0w_a$CDM model using different datasets.
We found that the GW data can effectively break the degeneracies between $H_0$, $\Omega_{\rm m}$, $w_0$, and $w_a$ in the BAO, SNe Ia, and BAO+SNe Ia datasets. Similarly, the addition of BAO and SNe Ia data can also improve the $H_0$ constraints by improving the constraints on $\Omega_{\rm m}$, $w_0$, and $w_a$. Compared with GW data only, the GW+BAO, GW+SNe Ia, and GW+BAO+SNe Ia data improve the $H_0$ constraint precisions by 17$\%$, 11$\%$, and 19$\%$, respectively.
Moreover, we found that the combination of BAO and SNe Ia can mutually break degeneracies and improve the constraints on $\Omega_{\rm m}$, $w_0$, and $w_a$.
Therefore, the GW+BAO+SNe Ia data can achieve joint constraints on $H_0$, $\Omega_{\rm m}$, $w_0$, and $w_a$. Specifically, using the GW+BAO+SNe Ia data, we obtain $H_0=74.8^{+6.3}_{-8.9}\ \ksm$, $\Omega_{\rm m}=0.320^{+0.015}_{-0.012}$, $w_0=-0.775^{+0.072}_{-0.074}$, and $w_a=-0.80\pm0.47$. This result deviates from $w=-1$ and supports the existence of dark energy evolution at a $\sim2\sigma$ confidence interval, with the inferred $H_0$ value being more consistent with the distance ladder result. To further explore the evolution behavior of $w$, we reconstructed $w(z)$ in the $w_0w_a$CDM model from the posterior samples of $w_0$ and $w_a$ using the GW+BAO+SNe Ia dataset, as shown in Fig.~\ref{fig:wz}. We find that within the 1$\sigma$ credible interval, $w(z)$ exhibits a mild phantom-crossing behavior, with the transition occurring at approximately $z\sim0.5$. However, within the 2$\sigma$ credible interval, $w(z)$ remains consistent with a cosmological constant as in the $\Lambda$CDM model.

When constraining cosmological parameters using GW standard sirens, the assumption of the GW population model can affect the inferred results. In this work, we adopt the Power Law + Peak model to describe the mass distribution of 42 BBH events, primarily because Ref.~\cite{LIGOScientific:2021aug} has revealed that the Power Law + Peak model is the most supported population model based on these BBH events. After the release of LVK O4a data, the most favored BBH population model became the Broken Power Law + 2 Peaks model \cite{Callister:2023tgi, LIGOScientific:2025pvj}, and the Power Law + Peak model has not been ruled out. Following Refs.~\cite{LIGOScientific:2021aug, LIGOScientific:2025jau}, we do not consider the evolution of the mass distribution with redshift, since based on current GW data, this evolution is not robustly supported \cite{Fishbach:2021mhp, vanSon:2021zpk, Heinzel:2024hva, Lalleman:2025xcs, Gennari:2025nho}. As mentioned in Sec.~\ref{sec:method}, we fixed the correlation parameter $\epsilon=1$ describing the relation between the galaxy luminosity and the probability of hosting GW sources, because the results of Refs.~\cite{LIGOScientific:2021aug, LIGOScientific:2025jau} have shown that the assumption of the galaxy luminosity weight has little influence on the cosmological parameter constraints. Compared with Ref.~\cite{LIGOScientific:2021aug}, we obtain the $H_0$ constraint precisions worse by 30--40$\%$ using the same GW data and assuming the same GW population models, mainly because we allow the BBH population parameters to vary and jointly constrain them together with the cosmological parameters. Our results using 47 GW events are generally consistent with Ref.~\cite{Gray:2023wgj}.

\section{Conclusions}\label{sec:conclusions}

The Hubble tension in the $\Lambda$CDM model has reached the $\sim$6$\sigma$ level, and no known observational systematic errors can account for such a significant discrepancy. Recently released DESI DR2 BAO data, when combined with the Planck CMB measurements, show a $\sim$3$\sigma$ preference for a dynamical dark energy model with a phantom-crossing behavior, which further exacerbates the Hubble tension. Since dark energy dominates the cosmic expansion in the late universe, it is essential to explore its properties using only late-universe observations. In this work, we investigate the role of a novel combination of late-universe probes, GW standard sirens, BAO, and SNe Ia, in cosmological parameter constraints.

We employ 47 GW standard sirens from GWTC-3, together with the DESI DR2 BAO and the DESY5 SNe Ia. For GW events without EM counterparts, we infer their redshifts by cross-matching with the GLADE+ galaxy catalog and applying the GW population model, jointly constraining the GW population model parameters and cosmological parameters in the analysis. The BAO sound horizon $r_{\rm d}$ and the SNe Ia absolute magnitude $M_B$ are also free parameters and fitted simultaneously with the cosmological parameters. In the $\Lambda$CDM framework, we obtain $H_0=76.1^{+6.1}_{-11.5}\ \ksm$ and $\Omega_{\rm m}=0.310\pm0.008$. In the $w_0w_a$CDM model, we obtain $H_0=74.8^{+6.3}_{-8.9}\ \ksm$, $\Omega_{\rm m}=0.320^{+0.015}_{-0.012}$, $w_0=-0.775^{+0.072}_{-0.074}$, and $w_a=-0.80\pm0.47$, indicating that the dark energy exhibits a phantom-crossing at $z\sim0.5$ in the 1$\sigma$ level, with the inferred $H_0$ value being more consistent with the distance ladder measurement.

Our results demonstrate that GW standard sirens provide an independent $H_0$ measurement, effectively breaking the degeneracies between $H_0$ and $r_{\rm d}$ in BAO and between $H_0$ and $M_B$ in SNe Ia. Moreover, the combination of SNe Ia and BAO data can break parameter degeneracies with each other, yielding tight constraints on $\Omega_{\rm m}$ and dark-energy EoS parameters. Therefore, the combination of GW, BAO, and SNe Ia enables simultaneous constraints on $H_0$, $\Omega_{\rm m}$, and dark energy EoS parameters, offering a unique late-universe perspective on the cosmic expansion and the nature of dark energy.

\begin{acknowledgments}
We are grateful to Peng-Ju Wu, Shang-Jie Jin, and Tonghua Liu for fruitful discussions. This work was supported by the National Natural Science Foundation of China (Grants Nos. 12533001, 12473001, 12575049, 12205039, and 12305058), the National SKA Program of China (Grants Nos. 2022SKA0110200 and 2022SKA0110203), the China Manned Space Program (Grant No. CMS-CSST-2025-A02), the 111 Project (Grant No. B16009), and the Natural Science Foundation of Hainan Province (Grant No. 424QN215).

\section*{Conflict of Interest}
The authors declare that they have no conflict of interest.

\end{acknowledgments}

\bibliography{main}

\end{document}